\documentstyle[12pt]{article}
\begin{document}
\title{{ Double $Q^2$-rescaling model and the nuclear effect of the parton
 distribution functions}
\thanks{The project  supported  in  part  by National Natural Science
foundation of China, Doctoral Program Foundation of Institution of Higher Education
of China, and  Hebei  Province   Natural Science  Foundation and Hebei
Province Education Committee of China.}}
\author{{Zhenmin He$^{1, 2}$  Xiaoxia Yao$^{1, 2, 3}$  Chungui Duan$^{1,2}$}\\
{Guanglie Li$^{1,3}$  Hongan Peng$^{1, 4}$}\\
{\small 1.CCAST(World Laboratory) P.O.Box 8730, Beijing, 100080, P.R.China}\\
{\small 2.Physics Department, Hebei Teachers' University, Shijiazhuang, 050016,P.R.China}\\
{\small 3.Institute of High Energy Physics, Academia Sinica, Beijing, 100039, P.R.China}\\
{\small 4.Physics Department, Peking University, Beijing,100871, P.R.China}}
\date{}
\maketitle
\vskip 2cm
\begin{abstract}
\baselineskip 9mm
In order to overcome the shortcoming of nonconservation of nuclear momentum
existing in the original $Q^2$-rescaling model(O$Q^2$RM) and avoid introducing nuclear shadowing
factor, we proposed a double $Q^2$-rescaling model(D$Q^2$RM) for the parton distributions of
the bound nucleon. Using the experimental data of lepton-nucleus deep inelastic
scattering(DIS) and the condition of the nuclear momentum
conservation, the $Q^2$-rescaling parameters of various partons
 for Sn, Fe, Ca and C nuclei are determined.
The rescaling parameters of valence quark distributions are larger than unity
and graduauly increases with atomic number A, on the contrary, the rescaling
parameters of sea quark distributions and gluon distributions are smaller than
unity, and slowly decreases with A. By using this model, the experimental
data of the DIS process, the nuclear Drell-Yan process and $J/\psi$ photoproduction
 process are consistently and quite satisfactorily explained.
\end{abstract}
\vskip 2cm
\maketitle
%\begin{document}
\baselineskip 9mm
\section*{1.Introduction}
{\hskip 1cm} In 1982, the European Muon Collaboration at CERN discovered in the
muon-nucleus deep inelastic scattering(DIS) that the structure functions
$ F_2$  of
bound nucleon is quite different from that of free nucleon within the region
$0.1\leq x \leq 0.7$$ ^{[1]}$, this nuclear effect was thus called the EMC effect.
Since the discovery of the EMC effect, physicists have proposed several
models and made out more or less satisfactory explanation for it $^{[2]}$.
But, for the shadowing effect emerged within very small x region in DIS
process $^{[3]}$ and the A-dependence of the differential cross section of h-A
Drell-Yan process $^{[4]}$, only the constituent quark model, $Q^2$-
rescaling model $^{[5]}$ and the extended x-rescaling model $^{[6]}$, after
introducing the nuclear shadowing factor, can make out satisfactory
 explanation. In addition, only the three models mentioned above have made out
  quantitative description for $J/\psi$ photoproduction on the nuclear
 target.\\

{\hskip 1cm} In the $Q^2$-rescaling model with shadowing factor, except for the
 parameter relative to shadowing,for the distribution functions of valence
 quark, sea quark and gluon within the bound nucleon, a common $Q^2$-rescaling
 parameters $\xi(A)$ was employed. Although the model can explain several
 kinds experimental processes mentioned above, but as pointed by LI Guanglie et.
 al.$^{[6]}$ it leads to non-conservation of nuclear momentum, that is, 
 the momentum of a nucleus is no longer equal to the sum of momenta of  valence
 quarks, sea quarks and gluons including in the nucleus. They thought that for the
valence quark distribution and sea quark distribution, two different $Q^2$-rescaling
parameters should be employed.\\

{\hskip 1cm} As a phenomenological approach to investigate the parton distribution
functions of a bound nucleon, here we presented a model, in
which three $Q^2$-rescaling parameters were employed for the distributions of
valence quark, sea quark and gluon respectively, and the artificial shadowing
factor was not used. Due to restriction of nuclear momentum conservation,
only two among the three parameters are independent, so we call this model the double $Q^2$-rescaling model(D$Q^2$RM).\\

{\hskip 1cm} From large amounts of available experimental data, it can be seen that
 for the both of free nucleon and bound nucleon, the x distribution and its evolution
 with $Q^2$ of valence quarks are quite different from those of
sea quarks and gluons, The distribution of valence quarks is mainly restricted
within the region of $x>0.1$, the " hardness" of the distribution allows one
to imagine that the valence quarks are confined within a bag with a distinct
boundary and the behavior of its evolution is described by GLAP equation
$^{[7]}$. The EMC effect mentioned above mainly reflects the change of valence
quark distribution of bound nucleon as compared with free nucleon. For the
distributions of sea quark and gluon,however, the case is much complex. 
The distribution of sea quark and gluon mainly restricted within the region of
$x<0.1$, and quickly increases with x decreasing. The case is more obvious
for the gluon distribution. Therefore, they play dominant role in the region of
very small x$^{[8]}$. As well known, the longitudinal size of parton with
momentum fraction x is $\Delta z=\frac{1}{xp}$, p is the momentum of nucleon
in a certain frame of reference with high speed, then usual picture that partons
are confined in a colour bag becomes indistinct. In relation with the case, soft partons with very small x
value from adjacent nucleons lying in the same longitudinal line within
nucleus may penetrate and overlap each other. The smaller x value ,and the
larger the atomic number A, the more outstanding
the phenomena. Then, recombination or
fusion between two or several partons from different nucleons could occur.
The partons with very small x recombine and form into a parton with slightly larger x,
and this process naturally leads to the shadowing in very small x region and
antishadowing in slightly larger x region, thereby, after taking into
account the recombination, the artificial shadowing
factor is not needed.\\

{\hskip 1cm} Based on above mentioned understanding, we present D$Q^2$RM. Using this model, the
influence of nuclear environment on the parton distributions and the experimental
data on l-A DIS process, h-A Drell-Yan process and $\gamma^*$-A $J/\psi$
production can be consistently and satisfactorily explained. In contrast with the rescaling parameter $\xi_V(A)$ of valence quark
distribution is larger than unity and slowly increases with A increasing
, a characteristic feature of our model is that both the rescaling parameters
$\xi_S(A)$ of sea quark distribution and $\xi_G(A)$ of gluon distribution are
smaller than unity and slowly decrease with A increasing. This feature just
 right meets the needs of the $Q^2$ evolution picture of parton distributions
in small x region of bound nucleon and nuclear momentum conservation. We shall
present experimental support for this feature  in the section 3.\\

\vskip 1cm
\section*{2.The double $Q^2$-rescaling model \\
 and relative formulas}
{\hskip 1cm} For a nucleus A, our model assumes three
different $Q^2$ -rescaling parameters for the distributions of valence quarks,
sea quarks and gluons in the bound nucleon respectively:
$$q_{Vi}^A(x,Q^2)=q_{Vi}^N(x,\xi_V^AQ^2),$$
$$q_{Si}^A(x,Q^2)=q_{Si}^N(x,\xi_S^AQ^2),$$
$${\bar q}_{Si}^A(x,Q^2)={\bar q}_{Si}^N(x,\xi_S^AQ^2),$$
\begin{equation}
 G^A(x,Q^2)=G^N(x,\xi_G^AQ^2).
\end{equation}
where i denotes the flavor of quark. That is to say, the distributions of
various partons of bound nucleon may be obtained by $Q^2$ rescaling for the
distributions of corresponding partons of free nucleon. For all of nuclei and any
$Q^2$ values, the momentum conservation condition
\begin{equation}
\int\limits_{0}^{1}xdx{\{\sum\limits_{i}[q_{Vi}^{A}(x,Q^2)+q_{Si}^{A}(x,Q^2)+{\bar q}_{Si}^{A}(x,Q^2)]+G^{A}(x,Q^2)\}}=1.
\end{equation}
should be respected. Because of the restriction of formula(2), only two among
three $Q^2$-rescaling parameters are independent, so this model is called the
double $Q^2$-rescaling model.\\

\mbox{}\hspace{1cm} In the D$Q^2$RM, the structure function
and gluon distribution function of bound nucleon respectively is
\begin{equation}
F^A_2(x,Q^2)=\sum\limits_{i}e_{i}^{2}x[q_{Vi}^{N}(x,\xi_{V}^{A}Q^2)
+q_{Si}^{N}(x,\xi_{S}^{A}Q^2)+{\bar q}_{Si}^{N}(x,\xi_{S}^{A}Q^2)]
\end{equation}
and
\begin{equation}
xG^A(x,Q^2)=xG^N(x,\xi_{G}^{A}Q^2),
\end{equation}
$F_{2}^{A}(x,Q^2)$ denotes the average nucleon structure function of a ideal
nucleus with the equal number of protons and neutrons(N=Z=$\frac{1}{2}$A).\\

\mbox{}\hspace{1cm} The nuclear effect of the nucleon structure function and
gluon distribution may be represented by following ratios:
\begin{equation}
R^{A/D}(x,Q^2)=\frac{F_{2}^{A}(x,Q^2)}{F_{2}^{D}(x,Q^2)},
\end{equation}
\begin{equation}
R_{G}^{A/D}(x,Q^2)=\frac{G^{A}(x,Q^2)}{G^{D}(x,Q^2)},
\end{equation}
usually, these ratios calculated by a certain theoretical model are used to
compare with the experimental data.\\

{\hskip 1cm} For the Drell-Yan  process in p-A collision,
 the nuclear effect is
represented by the ratio
\begin{equation}
T^{A/N}(x_{t},Q^2)=\frac{\int\frac{d^{2}\sigma^{p-A}(x,x_t,Q^2)}{dxdx_t}dx}
{\int\frac{d^{2}\sigma^{p-N}(x,x_t,Q^2)}{dxdx_t}dx},
\end{equation}
where
$$\frac{d^{2}\sigma^{p-A}(x,x_t,Q^2)}{dxdx_t}
=\frac{1}{3}\frac{4\pi\alpha^2}
{3M^2}\sum\limits_{i}e_{i}^{2}\{[q_{Vi}^{p}(x,Q^2)+q_{Si}^{p}(x,Q^2)]
{\bar q}_{Si}^{A}(x_t,Q^2) $$
\begin{equation}
+{\bar q}_{Si}^{p}(x,Q^2)[q_{Vi}^{A}(x_t,Q^2)+q_{Si}^{A}(x_t,Q^2)]\},
\end{equation}
$$\frac{d^{2}\sigma^{p-N}(x,x_t,Q^2)}{dxdx_t}
=\frac{1}{3}\frac{4\pi\alpha^2}
{3M^2}\sum\limits_{i}e_{i}^{2}\{[q_{Vi}^{p}(x,Q^2)+q_{Si}^{p}(x,Q^2)]
{\bar q}_{Si}^{N}(x_t,Q^2) $$
\begin{equation}
+{\bar q}_{Si}^{p}(x,Q^2)[q_{Vi}^{N}(x_t,Q^2)+q_{Si}^{N}(x_t,Q^2)]\},
\end{equation}
where $M^2$ denotes the invarint mass of lepton-pair($\mu{\bar \mu}$) produced in
the process.\\

\mbox{}\hspace{1cm}Three $Q^2$-rescaling parameters $\xi_{V}, \xi_{S}$ and$
 \xi_{G}$
are determined by the following method: firstly, we choose suitable $\xi_{V}$ and$
\xi_{S}$ to explain the nuclear effect in l-A DIS process; secondly, we describe
p-A Drell-Yan process with the same values of $\xi_{V}$
 and $\xi_{S}$ to test them,
and finally determine $\xi_{G}$ using nuclear momentum conservation and explain
the nuclear effect of $J/\psi$ photoproduction process at nuclear target.\\

\mbox{}\hspace{1cm} For the parton distribution function of free nucleon, we adopt
the parton momentum distributions given by M.Gluck, E.Reya and
 A.Vogt(GRV)$^{[9]}$,
because this formalism allows to be evoluted inversely to $Q^2\approx1.0GeV^2$.
This is quite important point for getting the distribution functions of sea
quark and gluon of bound nucleon by formula(1) at $\xi_{S}^{A}<1$ and
$\xi_{G}^{A}<1$.\\

\mbox{}\hspace{1cm} At present, the statistics and the precision of experimental
data on l-A DIS process, h-A Drell-Yan process and $\gamma^{*}-A$
J/$\psi$ production
process is insufficient to test very precise theoretical calculation, so that in
our calculation, only the leading order approximation on the formulas of hard
subprocesses and parton distribution functions are considered.\\

\mbox{}\hspace{1cm} The $Q^2$-rescaling parameters $\xi_{V}^{A}$
,$\xi_{S}^{A}$ and $\xi_{G}^{A}$
 on nuclei $C^{12}$, $Ca^{40}$, $Fe^{56}$
  and $Sn^{119}$ obtained by
above method from the data of references[3, 4, 10] are listed in the table 1. It is
worth to note that, all of $\xi_{V}^{A}$ are larger than unity and increase
with A increasing, but all of $\xi_{S}^{A}$ and
$\xi_{G}^{A}$ are smaller than
unity and decrease with A increasing. In the next section , we shall present
a experimental evidence supporting this result.\\

\begin{center}{Table.1}\end{center}
\[
\tabcolsep 0.5in
\begin{tabular}{|c|c|c|c|}\hline
       & $\xi_V$  &$ \xi_S$  & $\xi_G$ \\ \hline
 C     & 1.3    & 0.7    & 0.86   \\ \hline
 Ca    & 1.35   & 0.67   & 0.81   \\ \hline
 Fe    & 1.41   & 0.62   & 0.76   \\ \hline
 Sn    & 1.57   & 0.45   & 0.58   \\ \hline
 \end{tabular}
\]

\vskip 1cm
\baselineskip 9mm
\section*{3. Results and discussion }
\mbox{}\hspace {1cm} Using the values of $Q^2$-rescaling parameters listed in
the table 1 and formulas (1), (3) and (4), in which LO parton momentum
distributions of free nucleon given by GRV are adopted$^{[9]}$, we obtained various
parton distribution functions and average structure function $F_2$ of bound
nucleon in  nuclei $C^{12}, Ca^{40}, Fe^{56}$ and
$Sn^{119}$. Furthermore, using
formulas (5)-(9), the theoretical predictions on $R^{A/D}(x,Q^2)$
,$ T^{A/N}(x_t, Q^2)$ and $R_G^{A/D}(x, Q^2)$ by the double $Q^2$-rescaling 
model are calculated.\\
 
{\hskip 1cm} Theortical curves of $R^{A/D}(x,Q^2)$ on l-A DIS processes for 
nuclei $C^{12}$, $Ca^{40}$, $Fe^{56}$, $Sn^{119}$ within the $Q^{2}$ range given in reference[3] are separately plotted in Fig.1, where corresponding experimental data[3] are shown. It can be seen that by our model, only taking different parameters for 
valence quark distribution and sea quark distribution, without introducing the 
the shadowing factor, the experimental data of nuclear effect in l-A DIS process
can be explained quite well in the almost whole x region.\\

{\hskip 1cm} Theoretical curves of $T^{A/D}(x_t, Q^2)$ on p-A Drell-Yan process
for nuclei $C^{12}$, $Ca^{40}$ and $Fe^{56}$ within region of
 $0.025\leq x_t \leq 0.30$ and $4\leq M_{l\bar l} \leq 9GeV$
  and at $E_{CM}=40GeV$ are separately plotted in Fig.2a-c,
 where corresponding experimental data$^{[4]}$ are shown. We can see
that our results are  in qualitative agreement with the data, and especially in
the region of $x_t\leq 0.15$, the agreement is quite satisfactory. In addition,
within the region $0.15 \leq x_t \leq0.30$, the agreement
 case for heavier nucleus
is clearly better than  for lighter one. Furthermore comparing our results which 
the prediction given by original $Q^2$-rescaling model, it can be seen that
the D$Q^2$RM overcame the shortcoming that the tendency
of the curve given by original $Q^2$-rescaling model is wholly different from
the data.\\

{\hskip 1cm} Theoretical curve of $R_{G}^{Sn/C}(x,Q^2)$ (the ratio of gluon
distribution function of nucleus Sn to one of nucleus C) is plotted in Fig.3,
 where the experimental data$^{[10]}$ of the ratio measured in $\gamma^{*}$+A
$\rightarrow$ J/$\psi+X$ process are shown, the result calculated by original
$Q^2$-rescaling model is also plotted in the Fig.3.   Although there are quite large
errors in the experimental data, it can still be seen that the prediction given
by our model is clearly better than  O$Q^2$RM.\\

{\hskip 1cm} Curves of $R_{G}^{A/D}(x,Q^2)$  from D$Q^2$RM within region $0.1\leq
x\leq 0.2$ for nuclei C,Ca,Fe and Sn are plotted in Fig.4. It can be seen that slope
of these curves slowly increasing with A increasing. This case reflects that the
influence of gluon recombination is enhanced with A increasing.\\

{\hskip 1cm} In the following paragraphs, we shall discuss the problem
of the rescaling parameters $\xi_{V}>1$,$\xi_{S}<1$ and $\xi_{G}<1$.\\

{\hskip 1cm} From above comparision of the calculating results obtained by our
model with the experimental data we can see that our model has certain rationality,
and specific values of $\xi_{V}, \xi_{S}$ and
$\xi_{G}$ listed in the table 1 are
supported by the relevant experiments .\\

{\hskip 1cm} Firstly, in our model, $\xi_{V}>1$. This is the same as the O$Q^2$RM. The latter model can well explain the nuclear effect of
the DIS process in the region $0.2<x<0.7$, where the momentum distribution of
nucleus is mainly determined by valence quark, so the value of $\xi_{V}$ in our
model should roughly be consistent with the latter. In physics, $\xi_V>1$ can be explained by the swelling of nucleon. When the nucleon is swollen, according
to uncertainty principle, the momentum distribution of valence quarks moves to
small x region, i.e. the momentum distribution of valence quarks was softened.\\

{\hskip 1cm} Secondly, because of the following reasons , we assume that $\xi_
{S}<1$. The first reason is that  the O$Q^2$RM can not explain
the experimental data of nuclear effect of the l-A DIS process and nuclear
Drell-Yan process in small x region, where the nuclear effect is mainly
determined by the change of sea quark distribution, so we need to reconsider the
$Q^2$-rescaling parameter of sea quark. The second reason
 is that some authors of
this paper calculated the values of the nuclear effect function $R^{A}_{S}
(x, Q^2)$ of sea quark distribution by using the experimental data of l-A DIS
process and nuclear Drell-Yan process$^{[11]}$ and found that in region
$x\leq 0.2$ $R^{A}_{S}(x,Q^2)$ increases from the value smaller than  unity to
the value slightly larger than unity with x increasing, and the larger the
number A, the smaller the value of $R^A_{S}(x,Q^2)$. If we use $Q^2$-rescaling
model to describe this change of sea quark distribution influenced by nuclear
environment, it must demands $\xi_{S}<1$, and the larger the A, the smaller
the $\xi_{S}$.\\

{\hskip 1cm} Thirdly, we shall present another experimental surport on $\xi_G<1$.
It is well known that the gluon distribution of nucleon is the most important in
the small x region, and in the region of $x<10^{-2}$, gluon distribution
 almost saturates the parton distributions of nucleon.
 In the case, structure function $F_{2}(x,Q^2)$ is fully determined by the gluon
 distribution, and it can be obtained from PQCD$^{[12]}$
%\begin{equation}
\begin{eqnarray*}
 F_2(x,Q^2)&=&\sum\limits_{i}e_{i}^{2}\int\limits_{x}^{1}dz\int\limits^{Q^2}
\frac{dk^2}{k^2}P_{q_iG}(\frac{x}{z})\frac{\alpha_s(k^2)}{4\pi}
\frac{\partial [zG(z,k^2)]}{\partial(\ln k^2)}\\
          &\approx&\sum\limits_ie_{i}^{2}\frac{\alpha_s(Q^2)}{4\pi}
          \int\limits_{x}^{1}P_{q_iG}(\frac{x}{z})zG(z,Q^2)dz,\\
\end{eqnarray*}
\begin{equation}
\end{equation}
where $G \rightarrow$
$q_{i}$ splitting function $P_{q_iG}(x)={1\over 2}[x^2+(1
-x)^2]$. Due to the small the x value, the more important the gluon distribution,
therefore equation (10) can be approximately expressed as
\begin{equation}
 F_2(x,Q^2)\sim\frac{1}{8\pi}\sum\limits_ie_{i}^{2}\alpha_s(Q^2)G(x,Q^2)\Delta x.
\end{equation}
\baselineskip 9mm
Recently, NA37/NMC collaboration at CERN measured the ratio $F_2^{Sn}/F_2^{C}$
in the region of $0.0125 \leq x \leq0.55$ and $1\leq Q^2
\leq 10^2GeV^2$$^{[13]}$
their result shows that in the region of $x<3\times 10^{-2}$,
$F_2^{Sn}(x,Q^2)/F_2^{C}
(x,Q^2)$ is clearly less than unity, and this result means $G^{Sn}(x,Q^2)<G^{C}
(x,Q^2)$ in the region. If we use $Q^2$-rescaling approach to get $G^{Sn}(x,Q^2)$
from $G^C(x,Q^2)$ and assume
\begin{equation}
G^{Sn}(x,Q^2)=G^C(x,\xi^{Sn/C}Q^2),
\end{equation}
where $\xi^{Sn/C}$ should be smaller than unity ,that is to say, the gluon
distribution function of a heavier nucleus may be obtained from one of a
slighter nucleus by $Q^2$ rescaling approach and using $Q^2$ rescaling parameter
smaller than unity. Therefore, we may conjecture that the gluon distribution
functions of all nuclei may be obtained from one of free nucleon or deuterium
by this approach and using $\xi_G^{A}$ smaller than unity, moreover, the
values of
$\xi_G^{A}$ slowly decreases with atomic number A increasing.\\

{\hskip 1cm} In summary, we presented D$Q^2$RM, in which
different rescaling parameters $\xi_V$,$\xi_S$
 and $\xi_G$ are assumed, and the
shadowing factor is not introduced. By this model, the experimental data of the
l-A DIS process, the nuclear Drell-Yan process and $\gamma^{*}-A$
 J/$\psi$
production process were  consistently and quite satisfactorily explained. But,
 this model is a initial and phenomenological model, on the one hand, it needs
 to be further improved as more precise experimental data presented, on the
 other hand, its physical foundation still needs to be explored more
 elaboratively.\\

{\hskip 1cm} The authors are grateful to Dr.Cong-Feng QIAO for useful help
in computing work.\\

\newpage
%\begin{center}{FIGURE CAPTIONS}\end{center}
%\begin{enumerate}
%\item
%\end{enumerate}
\newpage
\centerline{\bf Figure Captions}
{\hskip 1cm}{\bf Fig.1} The ratio $R^{A/D}(x,Q^2)$  for 
nuclei $C^{12}$, $Ca^{40}$, $Fe^{56}$, $Sn^{119}$ versus the momentum fraction x within the $Q^{2}$ range given in 
reference[3], where corresponding experimental data[3] are shown. The solid lines are the results of our model, the dashed lines are the results of the old $Q^2-$rescaling model(In the OQRM, the $Q^2$-rescaling parameters  $\xi$ for nuclei $C^{12}$, $Ca^{40}$, $Fe^{56}$, $Sn^{119}$ are 1.6, 1.86, 2.02, 2.24, respectively).\\

{\hskip 1cm}{\bf Fig.2}
 The ratio $T^{A/D}(x_t, Q^2)$
for nuclei $C^{12}$, $Ca^{40}$ and $Fe^{56}$ versus x  within region of $0.025\leq x_t \leq 0.30$ and
 $4\leq M_{l\bar l} \leq 9GeV$
  and at $E_{CM}=40GeV$, 
 where corresponding experimental data$^{[4]}$ are shown. The meaning of lines is the same as that in Fig.1.\\

{\hskip 1cm} {\bf Fig.3} The ratio  $R_{G}^{Sn/C}(x,Q^2)$ 
versus x, 
 where the experimental data$^{[10]}$  are shown. The meaning of lines is the same as that in Fig.1.\\

{\hskip 1cm} {\bf Fig.4} The ratio  $R_{G}^{A/D}(x,Q^2)$ 
  from D$Q^2$RM versus x  within region $0.1\leq
x\leq 0.2$ for nuclei $C^{12}$, $Ca^{40}$ and $Fe^{56}$. \\


\begin{thebibliography}{s2}
\bibitem{s1} EMC,J.J.Aubert et.al., Phys.Lett.,B123(1983)275.
\bibitem{s2} C.H.Llewellyn Smith, Phys.Lett.,B128(1983)107.\\
             M.Ericson,A.W.Thomas,Phys.Lett.,B128(1983)112.\\
             F.E.Close,R.G.Roberts and G.C.Rose, Phys.Lett.,B129(1983)346.\\
             R.L.Jaffe et.al., Phys.Lett.,B134(1984)449.\\
             F.E.Close et.al., Phys.Rev.,D31(1985)1004.
\bibitem{s3} NMC,P.Amaudruz,et.al., Z.Phys.,C51(1991)387;Nucl.Phys.,B441(1995)3.\\
             J.Ashman et.al., Phys.Lett.,B202(1988)603.
\bibitem{s4} D.M.Alde et.al., Phys.Rev.Lett.,64(1990)2479;66(1991)2285.\\
             M.R.Adams et.al., Phys.Lett.,B287(1992)375.
\bibitem{s5} M.Arneodo, Phys.Rep.,240(1994)301.
\bibitem{s6} G.L.Li,Z.J.Caoand C.S.Zhong,Nucl.Phys.,A509(1990)757.
\bibitem{s7} G.Altarelli and G.Parisi,Nucl.Phys,B126(1977)298,\\
             V.N.Gribov and L.N.Lipatov,Sov.J.Nucl.Phys.15(1972)78,\\
             Yu.L.Dokshitzev,Sov.J.Nucl.Phys.,73(1977)1216.
\bibitem{s8} HI Collab.,DESY 95-068.
\bibitem{s9} M.Gluck,E.Reya and A.Vogt,Z.Phys.,C67(1995)433.
\bibitem{s10} NMC,P.Amaudruz et.al.,Nucl.Phys.,B371(1992)553.
\bibitem{s11} Zhen-Min HE et.al.,Commun.Theor.Phys.,(to be published).
\bibitem{s12} G.Altareli,Phys.Rep.,81(1982)1.
\bibitem{s13} NMC Collab.,A.Mucklich,Proceeding of the Workshop on Deep
              Inelastic Scattering and QCD.(p.489),Paris,April.1995.
\end{thebibliography}
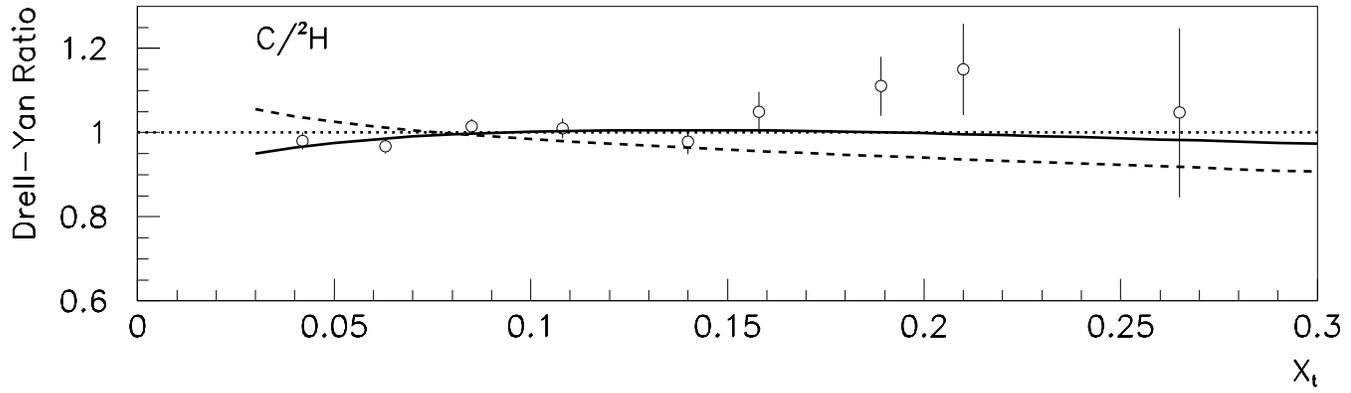
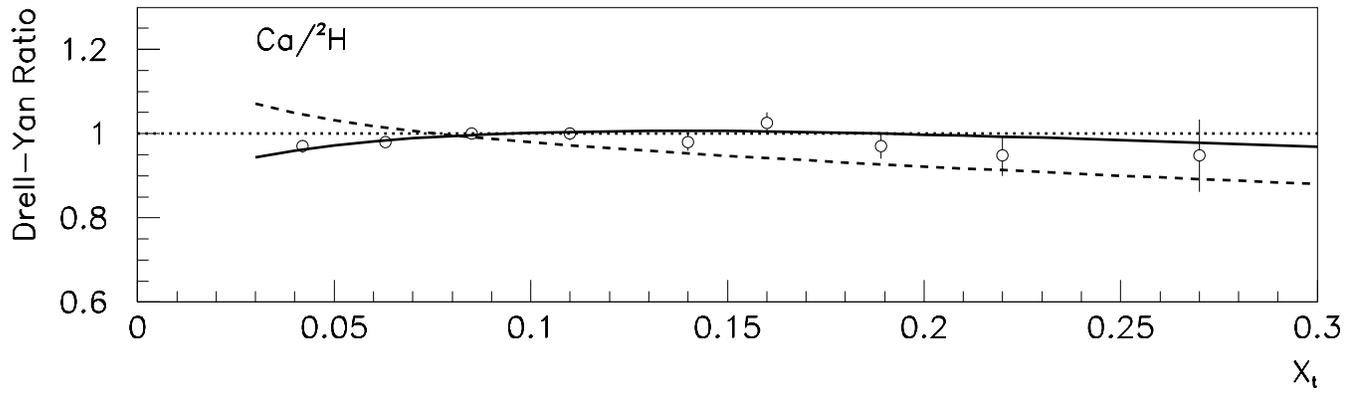
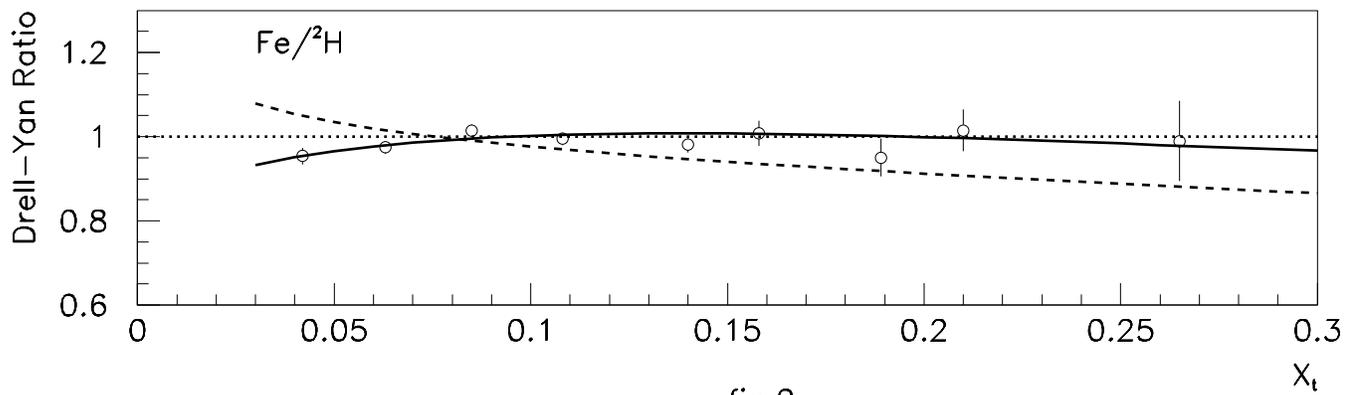
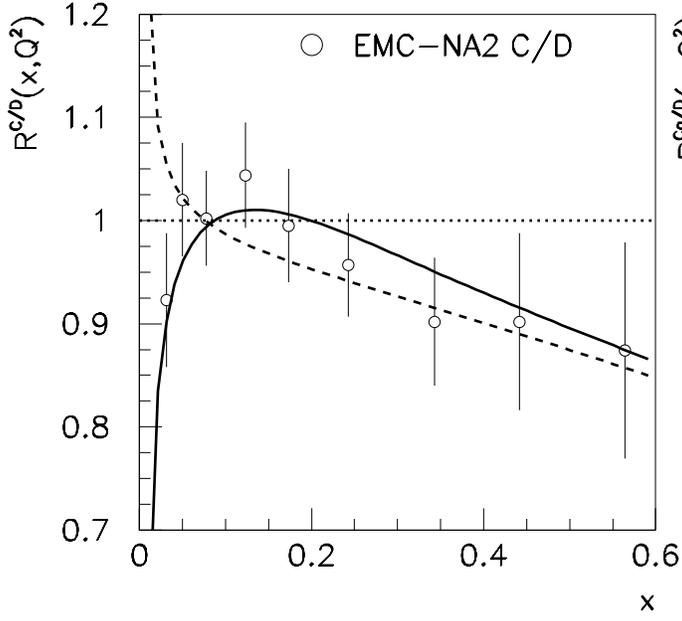
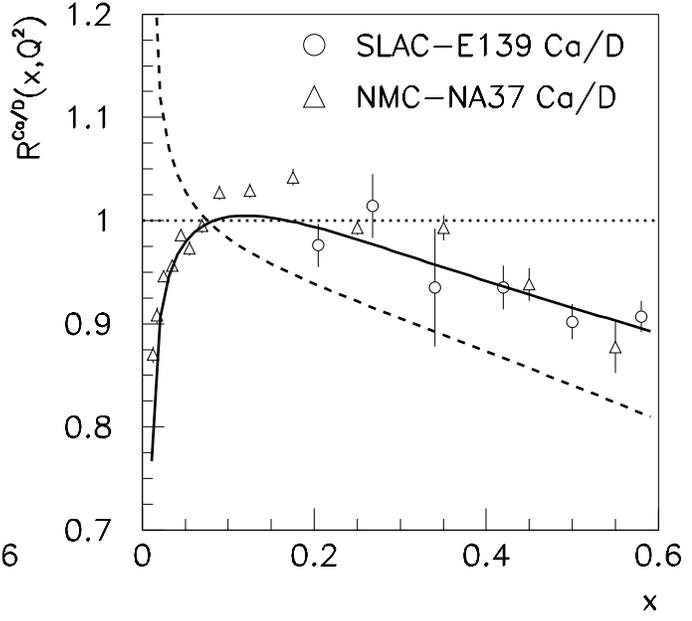
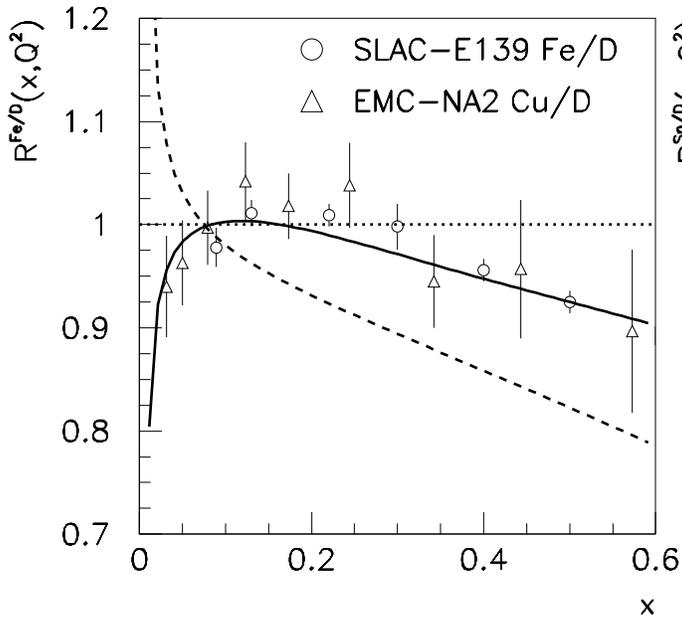
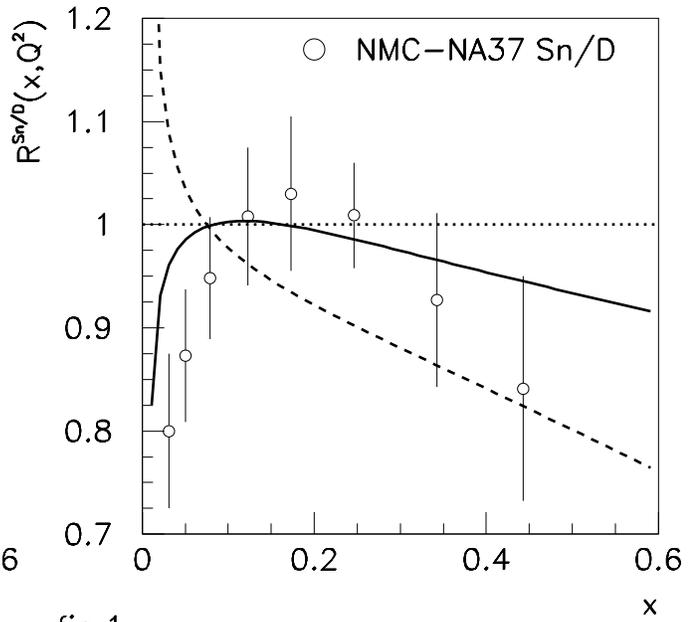
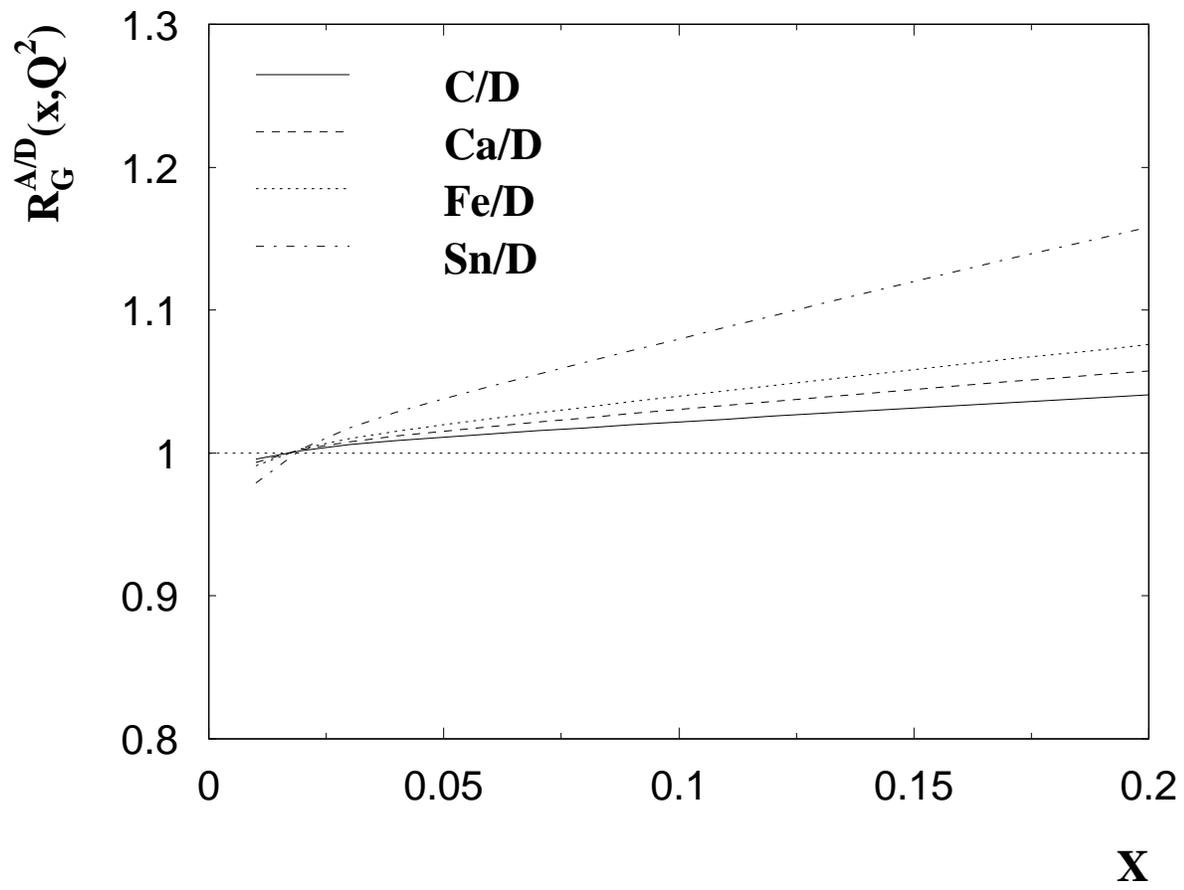
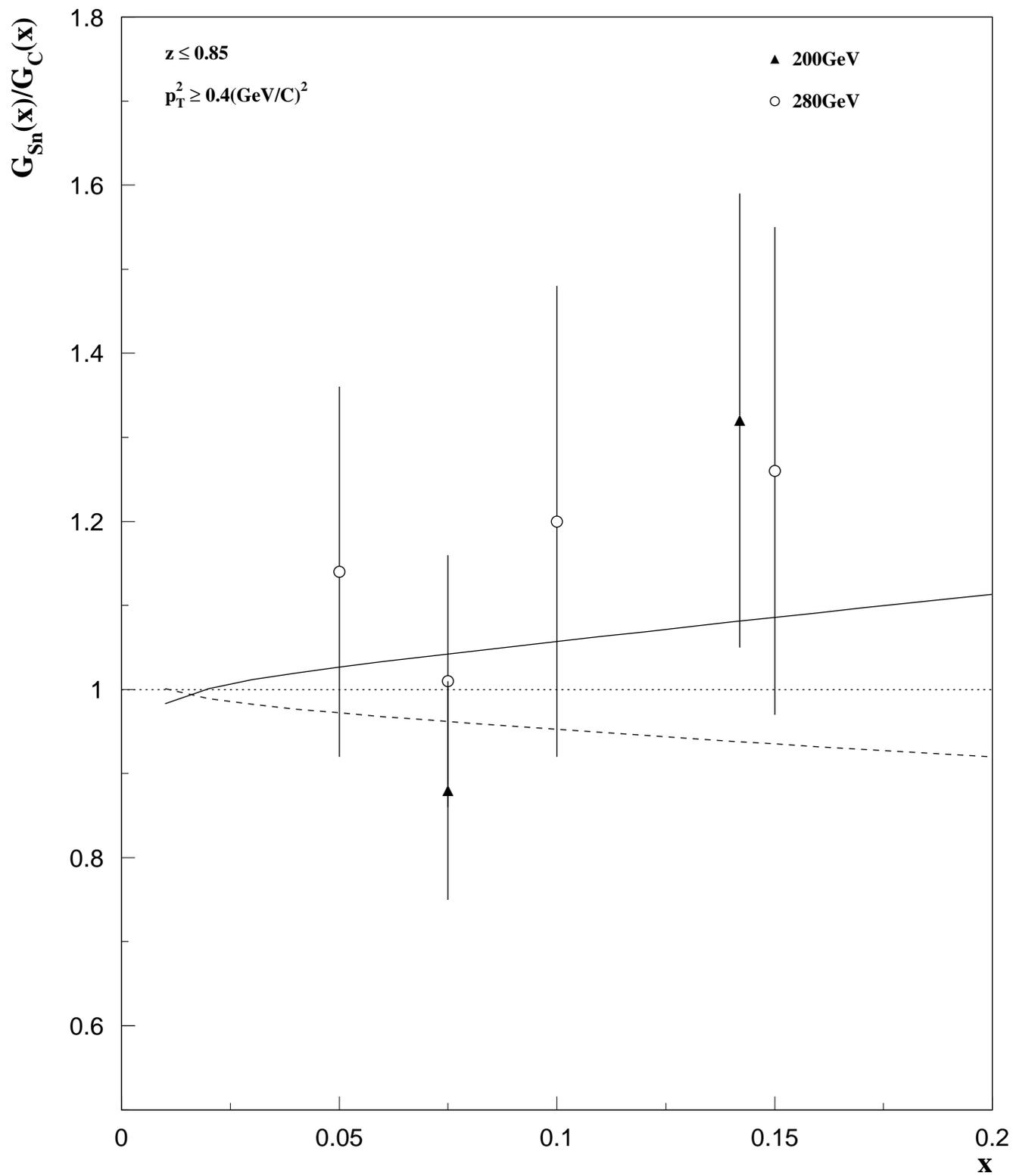
\end{document}